# Large-Area Emergency Lockdowns with Automated Driving Systems

Noah Goodall

Virginia Transportation Research Council, USA, noah.goodall@vdot.virginia.gov

**Abstract**
Region-wide restrictions on personal vehicle travel have a long history in the United States, from riot curfews in the late 1960s, to travel bans during snow events, to the 2013 shelter-in-place "lockdown" during the search for the perpetrator of the Boston Marathon bombing. Because lockdowns require tremendous resources to enforce, they are often limited in duration or scope. The introduction of automated driving systems may allow governments to quickly and cheaply effect large-area lockdowns by jamming wireless communications, spoofing road closures on digital maps, exploiting a vehicle's programming to obey all traffic control devices, or coordinating with vehicle developers. Future vehicles may lack conventional controls, rendering them undrivable by the public. As travel restrictions become easier to implement, governments may enforce them more frequently, over longer durations and wider areas. This article explores the practical, legal, and ethical implications of lockdowns when most driving is highly automated, and provides guidance for the development of lockdown policies.

**INTRODUCTION**
Travel restrictions have long been used by governments to control population movement during emergencies. These restrictions can take various forms. Individual roads, bridges, and tunnels may be closed to private vehicle traffic, as was done for all routes into Manhattan following the terrorist attacks of September 11, 2001 (*1*). Curfews may also be enforced, where pedestrian and vehicle traffic are prohibited during certain hours. Several jurisdictions have curfew laws for minors, and curfews in the United States have been enforced for all adults during emergencies such as the 1965 Watts riots in Los Angeles, California (*2*). Authorities may order citizens to "shelter-in-place" and remain in their homes or business during an emergency, although several states do not have specific statutes that allow mandatory shelter-in-place orders (*3*).

In recent years, the terms "shelter-in-place" and "lockdown" have gained prominence in emergency management. While often used interchangeably, these terms have distinct meanings in the United States. A lockdown typically refers to situations involving an active shooter (*4*), while shelter-in-place orders are generally issued for extreme weather events or potential chemical/radiation exposure (*5*).

Governments rarely enforce large-area lockdowns due to their unpopularity, cost, and enforcement challenges. Transportation agencies may need to deploy barriers, while police or military personnel must enforce closures. The cost of this additional personnel is in addition to the lost tax revenue from closing businesses. In the United States, lockdowns have generally required public support or a large police or National Guard presence, and often both.

These challenges are mitigated with the introduction of automated driving systems (ADS). SAE International defines an ADS as "the hardware and software that are collectively capable of performing the entire DDT [dynamic driving task] on a sustained basis" (*6*). More advanced ADS can perform the entire driving task without any input from a passenger and may thus operate without a human driver/passenger. These ADS, referred to as Levels 4 and 5, may or may not be equipped with conventional controls (*6*). Because these ADS would be completely reliant on their software and wireless communication systems to operate, governments could potentially restrict their movements simply by coordinating with a few ADS developers, fleet owners, or mapping service providers. Alternatively, authorities might be able to place low-cost barriers such as traffic cones and key locations and leverage the software's adherence to traffic laws. Enacting and enforcing lockdowns could become far more cost effective, leading to potential abuse by governments without a firm legal framework for their permissible usage.

This article considers the potential risks of mandatory shelter-in-place orders, referred to as "lockdowns," during emergencies in a future when automated driving systems lacking the controls to be operated by a human passenger are in widespread use.

The primary objective of this research is to examine the practical, legal, and ethical implications of large-area lockdowns in a future where automated driving systems are prevalent. Specifically, this paper aims to:

1. Analyze the characteristics and legality of historical lockdowns using the 2013 Boston Marathon bombing response as a case study.
2. Explore the potential methods by which governments could restrict ADS movements during emergencies.
3. Propose guidelines for the ethical initiation and enforcement of lockdowns in the era of automated driving.

As automated driving systems become more sophisticated and widespread, the ease with which governments can implement travel restrictions may lead to their more frequent use, potentially over longer durations and wider areas. This raises important questions about the balance between public safety and individual liberty, the role of private technology companies in emergency management, and the potential for abuse of these capabilities.

The remainder of this article is organized as follows. First, characteristics of lockdowns are described, with the 2013 voluntary lockdown in Boston, Massachusetts following a terrorist attack as a case study. Then, ways in which automated driving systems allow for inexpensive and quick enforcement of lockdowns are explored. Finally, guidelines for ethical initiation and enforcement of lockdowns are presented for policymakers, ADS developers, and emergency management professionals.

By addressing these issues, this paper contributes to the growing body of literature on the societal impacts of automated driving systems and provides valuable insights for transportation planners, emergency management officials, and policymakers as they prepare for a future where ADS play a central role in personal mobility.

**LOCKDOWN CHARACTERISTICS**

**Boston Case Study**
On April 15, 2013, at 2:49 p.m., two bombs detonated 12 seconds apart near the finish line of the still underway Boston Marathon, killing three spectators and injuring hundreds more. Two suspects were later identified from video surveillance. Before they could be apprehended, the suspects killed a police officer in nearby Cambridge and exchanged fire with police in Watertown, seven kilometers from the initial bombing. One suspect escaped from the shootout. By 2 a.m. the following morning, Boston-area residents began receiving automated phone calls and notes on their doors advising them to shelter-in-place (*7*). Shortly after 8 a.m. on April 19, Massachusetts Governor Deval Patrick requested that Boston-area residents stay indoors, not answer the door unless it is a police officer, stay away from windows, and not congregate outside (*8*). Police cordoned off a 20-block area in Watertown and conducted building-by-building searches for the remaining suspect. The shelter-in-place request was lifted shortly after 6 p.m. that day (*8*). Sometime later, a Watertown resident living outside the cordoned area found the injured suspect lying in a boat parked in his backyard. By 9 p.m. police had the suspect in custody (*9*).

During the 16-hour lockdown, public transit was shutdown, Amtrak passenger trains to and from New York were stopped and searched (*10*), taxis were ordered off the street for part of the day, professional baseball and hockey games were cancelled, and nearly all businesses, offices, and universities were closed (*8*). Reporters observed nearly empty streets throughout the day, a stark contrast to the usual Friday pedestrian and vehicle traffic (*8*).

The lockdown received widespread support in the Boston area. A survey of 500 Massachusetts residents conducted after the lockdown by the independent group MassINC that showed 91% of respondents approved of actions taken by the government, with 48% fearing the response would not go far enough to end terrorism (*11*). Media interviews of Boston residents also demonstrated nearly universal support of police actions (*12*). The elected representative of Massachusetts's 8th congressional district, whose constituents include Bostonians, reported receiving no complaints about the shelter-in-place request (*13*).

Outside of Boston, reactions were mixed. In a national survey conducted by *The Washington Post* during the lockdown, 48% of respondents worried the government would go too far in its investigation because of concerns about constitutional rights versus 41% who thought they wouldn't go far enough (*14*). Images of police in combat uniforms and body armor traveling on suburban streets in armored personnel carriers drew comparisons to a military occupation (*13*). Although the shelter-in-place request was voluntary, some commentators interpreted it as "strongly" suggested (*15*), with the governor's statement advising that residents "should not" answer their doors and noting that private businesses "will remain closed until further notice" (*8*). Others objected to the cost of the lockdown, estimated at between $250 to $330 million (*16*).

A lockdown of Boston's scale was unprecedented in the U.S. at the time (*8*). Major fugitive hunts in the aftermath of the Oklahoma City bombing in 1995, the 2002 mid-Atlantic sniper attacks, and the 1996 Atlanta Olympic bombing all avoided lockdowns. The only comparable terrorism-related lockdowns in recent history followed the attacks of September 11, 2001 where five percent of Manhattan closed to vehicle traffic for 30 days (*1*), and the 2015 Paris attacks when Brussels was locked down for five days while authorities searched for suspects (*17*).

**Legal Aspects**

In the United States, the rights of citizens to travel between individual states is generally protected by the Constitution (*18*). In an emergency, however, fundamental rights such as travel may be temporarily limited or suspended (*18*).

While travel between states is a protected right, many state and federal courts have reached differing conclusions regarding whether travel within a state is constitutionally protected (*18*). Although the Supreme Court has declined to decide the issue, several states' constitutions establish a state constitutional right to travel within the state (*18*). Curfew laws at the city or county level have been held as valid against constitutional objections, so long as they are enacted or promulgated during civil disorder or riots, and if reasonable notice is given (*19*). In Massachusetts, mayors and city managers are permitted to enact curfews in response to civil disorder or other dangers, provided two hours' notice is given (Mass. Gen. Laws Chap. 40 § 37A). For an extensive analysis of intrastate travel rights, see Wilhelm (*20*).

Internationally, freedom of movement is generally protected under Article 13 of the Universal Declaration of Human Rights (*21*), a document developed by the United Nations in 1948 with 48 countries voting for its adoption by the UN General Assembly.

**Costs**

Officials may be reluctant to lock down a city during an emergency due to the costs. The Boston region has a gross domestic product of almost $1 billion USD per day, but the lockdown affected only a portion of the region and much of the lost revenue was simply delayed; experts estimated the economic losses at between $240 to $333 million per day for Boston (*16*).

Cities may need additional police to enforce a lockdown, such as the 19,000 National Guard troops deployed to Boston during the lockdown (*22*). Boston has one sworn officer for every 317 residents (*23*). If working alternating 12-hour shifts, one officer would be on duty for every 634 residents. In contrast, the standard ratio of police officers to attendees for effective crowd control is 1:100 (*24*). As one-fifth of Boston's payroll is devoted to police (*25*), a six-fold increase in police presence for any significant length of time would be unsustainable.

**LOCKDOWNS AND AUTOMATED DRIVING SYSTEMS**
The Boston lockdown was voluntary and occurred in 2013 when all vehicles were manually driven by humans. The act of driving has become increasingly computerized in the years since. Anti-lock brakes, stability control, and cruise control have advanced into adaptive cruise control and lane-keeping functions on productions vehicles, automating the gas, brake, and steering functions in certain conditions. These features rely on advanced cameras, laser and radar sensors, and computerized control of steering and braking. While today's production vehicles can drive themselves in limited scenarios, e.g., self-parking [cite] and well-marked freeways [cite], automakers and technology firms have deployed vehicles capable of handling complex urban traffic in range of weather and light conditions. At least one system is currently operating as a driverless taxi service in four U.S. cities with over 7.1 million miles traveled (*26*).

The increasing automation of the act of driving may simplify the enforcement of a mandatory lockdown. There are several ways that a government can quickly and cheaply restrict the movements of vehicles with ADS. Three examples are discussed in this section: jamming wireless communications, altering digital maps, and exploiting vehicles' strict adherence to traffic laws.

Each of these methods can be overcome by the human driver taking manual control of the vehicle. Future vehicles, however, may lack conventional controls, or controls may be restricted in some way to prevent operation by unlicensed or impaired drivers. Even when controls are available, passengers who may have never driven before may find themselves unable to drive. In these situations, it may be possible for passengers to request a human driver in an off-site operations center to control the vehicle using wireless communications. This process is referred to as teleoperation (*27*), and has been demonstrated several times using existing cellular networks (*28*). Remote operators could potentially control the vehicles and ignore the restrictions, but they may be unwilling to assist, their employers may be unwilling, governments may block the necessary wireless communication, and there may not be an adequate number of remote operators to service all vehicles in a lockdown at once.

Each of the methods governments can use to restrict the movement of vehicles with ADS is explored in this section.

**Jamming Wireless Communication**
Most automated driving systems use occasional wireless communication via cellular networks, either to communicate with remote operators (*29*), receive data on road conditions (*29*), or updating the high-definition (HD) maps required for automated driving (*30*). In order to restrict automated driving, officials may attempt to disrupt wireless communications between vehicles and the infrastructure. This could be accomplished through deploying devices that deliberately interfere with radio or GPS signals, referred to as signal jammers (*31*, *32*). In the United States, the use of these technologies by state and local authorities are prohibited under the Communications Act of 1934. Federal agencies may jam signals after seeking a waiver, and sources have acknowledged using these devices during the 2009 presidential inauguration ceremony in Washington, DC to prevent remote controlled bomb attacks (*33*). Although state governments are currently prohibited from jamming signals, they have repeatedly requested exemptions (*34*).

**Altering Digital Maps**
If automated vehicles cannot be driven manually by their passengers, then governments wishing to restrict the movement of automated vehicles have several available strategies. Road closures could be entered into navigation map services such as Google Maps, Apple Maps, Waze, or other proprietary navigation maps used by ADS developers. This would prevent a vehicle's navigation system from using roads into and within a lockdown area.

Governments may only need to partner with a handful of companies to effect a lockdown using these methods. As of 2020, two companies, Alphabet (parent company of ,Google Maps and Waze) and Apple have near-complete control of the navigation software market (*35*). Similarly, two companies—Waymo and Cruise—constitute 99% of all driverless vehicle miles in California (*36*). This may not require explicit cooperation from mapping companies, as transportation agencies routinely send road closure information to mapping services automatically (*37*).

Most ADS also rely on high-definition (HD) maps of most roads, comparing the locations of various objects from a vehicle's lidar sensors with map data to distinguish between static and moving objects, and to geolocate a vehicle's position by triangulating the distance to buildings and signs (*30*, *38*). These HD maps are routinely updated using the lidar data collected by probe vehicles, and new map details such as locations of construction barriers may be shared among Avs in near-real time. These maps are considered so critical to automated driving that the Chinese Internet search company Baidu at one point considered shifting its focus to primarily making and distributing HD maps as a service for automated driving systems (*39*).

Governments could request that HD map providers "close" roads or digitally insert representations of physical barriers in the map in such a way that ADS programming will not permit them to drive onto these roads. A map provider, for example, could insert a "digital fence" that an ADS will not drive into, regardless of its own sensor readings.

Although these strategies may require cooperation from ADS developers or mapping software companies, such cooperation from private industry is not unprecedented, especially in a crisis. YouTube, Facebook, Twitter, and Reddit cooperated with police to remove a video of a mosque shooting in Christchurch, New Zealand within minutes of its posting (*40*).

**Exploiting ADS Adherence to Traffic Laws**
Authorities might also deploy low-cost physical barriers that ADS have been programmed to respect as road closure warnings. A single traffic cone or police tape, even without an officer present, might be sufficient to close a road. Automated driving systems have been vulnerable to these strategies in real-world deployments, in one instance being immobilized by a traffic cone placed on a vehicle's hood (*41*).

Much of the cost of enforcing a vehicle travel ban can be minimized by using automated enforcement tools. Police can track a vehicle's location, time of movement, movement over time *i.e.* tracking, speed, and unique identification using GPS, toll tag sensors, license plate readers, Wi-Fi/Bluetooth sensors, or some other type of vehicle-to-infrastructure communication such as dedicated short-range communications (DSRC) (*42*).

**Human Driving Difficult or Impossible**
Vehicles equipped with ADS may not have conventional controls, making them undrivable for their human passengers. Some examples today are low-speed shuttles, in operation in several cities (*43*). Many low-speed shuttles with ADS do not have conventional controls such as a steering wheel or pedals, but rather handheld devices similar to a video game controller to allow the backup driver, or "conductor," to control the vehicle when the automated driving system cannot (*44*).

In the United States, the National Highway Traffic Safety Administration (NHTSA) has begun the process of adapting its safety standards to permit vehicles with non-conventional controls (*45*, *46*). If future vehicles are able to drive consistently without passenger intervention, then the described low-cost efforts to enact lockdowns (spoofing, jamming, and altering maps) may become even more effective as passengers will have no way to override the ADS and operate the vehicle.

**Staff Cannot Support Remote Operation**
If the passengers cannot drive the vehicle, and the automated driving system is impacted by lockdown measures, a driver employed by a third party may control the vehicle from a remote location using wireless communications. This is generally referred to as "remote operation" where an operator may observe the vehicle's environment and select a path for the ADS to follow, or "teleoperation" where a remote operator takes full control of the vehicle and is able to perform the driving task. At least seven ADS developers have the ability to operate vehicles remotely in some capacity, while five other companies specialize in remote operation technologies for automated vehicles (*28*).

Several states, including Texas, Florida, and California, explicitly permit remote operation of vehicles (*47–49*). Among these, some states require that remote operators be available during testing of automated vehicles without an on-board human driver (*49*, *50*). However, no jurisdiction mandates a one-to-one ratio of remote operators to automated vehicles.

ADS developers may never employ a one-to-one operator-to-vehicle ratio, as the only cost advantages it offers over a human driver in the vehicle are lower wages due to outsourcing (*51*). Instead, remote operators may work in teams, responsible for monitoring several vehicles and taking control as requested either by the passengers or the automated driving system. Using this approach, a few remote drivers could be responsible for a large vehicle fleet, resulting in substantial cost savings.

It is unlikely that the supply of available remote drivers at any given time will be able to handle the demand should an entire city suddenly require them for all driving. The number of remote operators needed to manage a fleet can be estimated using queueing theory. Call centers with known call durations and volumes use the Erlang C formula to estimate staffing needs (*52*). The Erlang C formula estimates the probability $P_c$ that an incoming call cannot be immediately answered given the total staffing $N$ and call demand $A$ using the equation:

$$P_C(N, A) = \frac{\frac{A^N N}{N!(N-A)}}{\sum_{i=0}^{N-1} \frac{A^i}{i!} + \frac{A^N N}{N!(N-A)}},$$

where $A$ is the average number of calls $\lambda$ divided by average number of calls that can be answered by a single operator in the same amount of time $\mu$ given in the equation $A = \lambda/\mu$.

Using this approach and assuming 100% of current vehicle travel using ADS, Goodall (*51*) estimated the number of remote operators needed to support all United States driving. Assuming that an ADS requires assistance once per 397 hours of operation—the rate at which Waymo ADS disengaged in 2018 in California—and if each assistance requires an average of five minutes, then 19,072 remote operators working shifts could support all United States driving. According to the United States National Household Travel Survey, vehicles in the United States are in operation 101 billion hours per year (*53*). When divided by 8,760 hours in a year, this suggests that 12 million vehicles are on the road at any given moment, and 147,000 in Boston. If remote drivers were needed for all these vehicles simultaneously, there would not be adequate remote drivers (19,072) in the United States to handle the assistance requests (147,000).

**GUIDANCE CRITERIA**

As automated driving systems (ADS) become more prevalent, the potential for governments to implement large-area lockdowns more easily and frequently raises significant ethical and practical concerns. This section proposes a framework for developing guidance criteria to ensure that such lockdowns are implemented justly, effectively, and with due consideration for individual rights and societal needs.

In emergencies with incomplete information, authorities may be tempted to make tactical decisions that restrict liberty. While such restrictions are often unavoidable in limited contexts like traffic stops, decisions affecting an entire city should be made using political reasoning and with careful consideration of their broader implications. As Harvard professor Herman "Dutch" B. Leonard noted in his testimony before Congress regarding the Boston lockdown:

> "For example, the decision to issue a shelter-in-place request was appropriately framed as a policy issue by operational commanders and was put to political leaders for resolution, and this may provide a good illustration of the kind of process of issue identification and resolution that needs to be addressed in the [National Incident Management System (NIMS)]. It is imperative for NIMS to provide more guidance about the process by which tactical commanders should work in conjunction with an appropriate process for decision making by elected leaders. Both have important but different roles to play, and NIMS currently lacks systematic ways to help these two groups each to stay within their own designated 'lane.'" (*7*)

This observation underscores the need for a clear decision-making framework that involves both operational and political leadership when considering large-scale lockdowns.

When rights must be restricted during an emergency, authorities should adhere to the "least restrictive alternative" principle to maintain safety and security while minimizing infringement on individual liberties (*54*). In the context of ADS-enabled lockdowns, this means limiting the scope of travel restrictions to the smallest necessary geographic area, minimizing the duration of the lockdown, restricting the fewest possible modes of travel, and allowing for essential movement within the restricted area. By adhering to these principles, authorities can balance public safety needs with respect for individual freedom and minimize potential economic and social disruption.

Jennings and Arras (*54*) recommend that authorities pursue voluntary compliance rather than mandatory enforcement whenever possible. This approach is ethically superior and often more effective, assuming that the necessary behaviors can be achieved. To encourage voluntary compliance, authorities should provide clear, transparent communication about the reasons for the lockdown, offer rational persuasion, correct misinformation promptly, and consider public opinion while addressing concerns proactively. However, when individuals pose a serious threat to others by their unwillingness to comply with restrictions, there may be ethical justification for compelling compliance.

The unique capabilities of ADS present both opportunities and challenges for implementing lockdowns. Guidance criteria should address the use of digital map alterations and geofencing to restrict ADS movement, protocols for emergency override of ADS controls by authorized personnel, safeguards against potential abuse or hacking of ADS lockdown mechanisms, and provisions for essential services and emergency vehicle operations.

To ensure that ADS-enabled lockdowns are implemented ethically, authorities should respect the privacy and confidentiality of individuals affected by the restrictions, protect restricted individuals from undue social stigma and humiliation, ensure that adequate resources are available to enforce requirements fairly and humanely, provide clear criteria for initiating, maintaining, and lifting lockdowns, and establish oversight mechanisms to prevent abuse of emergency powers.

Guidance criteria should also address exceptions to lockdown orders, such as medical emergencies and access to healthcare, essential workers and critical infrastructure maintenance, caregiving responsibilities and family reunification, and individuals with special needs or vulnerabilities. To prevent the unnecessary prolongation of emergency measures, guidance should include regular review of the continued necessity of lockdown measures, clear criteria for lifting or modifying restrictions, and sunset provisions that automatically terminate emergency powers after a specified period unless explicitly renewed.

Given the potential for cross-border implications of ADS-enabled lockdowns, guidance criteria should address coordination with neighboring jurisdictions and countries, compliance with international human rights standards and freedom of movement agreements, and protocols for managing international travel and border crossings during lockdowns.

By developing comprehensive guidance criteria that address these various aspects of ADS-enabled lockdowns, authorities can ensure that emergency measures are implemented effectively, ethically, and with due respect for individual rights and societal needs. As ADS technology continues to evolve, these criteria should be regularly reviewed and updated to reflect new capabilities, challenges, and ethical considerations.

**CONCLUSIONS**

This paper has examined the potential implications of automated driving systems (ADS) on the implementation of large-area emergency lockdowns. Three key points were explored: the historical context and characteristics of lockdowns, the ways in which ADS could facilitate more efficient lockdown enforcement, and the ethical and legal challenges this capability presents.

The analysis suggests that ADS-enabled lockdowns could be implemented more quickly and cost-effectively than traditional methods, potentially leading to their more frequent use. This capability raises important questions about the balance between public safety and individual liberties, as well as the role of private technology companies in emergency management.

A framework for developing guidance criteria for ADS-enabled lockdowns was proposed, emphasizing the importance of using the least restrictive measures necessary, promoting voluntary compliance, ensuring transparency, and establishing clear oversight mechanisms. This framework is intended as a starting point for further discussion and refinement by legal scholars, ethicists, and policymakers.

The goal of this paper was not to solve these complex issues definitively, but to lay out the challenges and potential consequences of ADS-enabled lockdowns. Future research should focus on developing more detailed protocols for ADS-enabled lockdowns, exploring the technical feasibilities and limitations of proposed restriction methods, and conducting analyses of potential societal impacts.